\documentclass[prl,twocolumn,showpacs,preprintnumbers,amsmath,amssymb]{revtex4-1}

\usepackage{graphicx}% Include figure files
\usepackage{dcolumn}% Align table columns on decimal point
\usepackage{bm}% bold math

\newcommand{\subel}[1]{{\mbox{\tiny #1}}}
\renewcommand{\vec}[1]{\mathbf{#1}}

\usepackage[usenames,dvipsnames,svgnames,table]{xcolor}

\newcounter{todoc}

\usepackage[active]{srcltx}
\begin{document}

%\preprint{APS/123-QED}

\title{Finite temperature magnetism of FeRh compounds}
\author{S.\ Polesya} 
\email{Svitlana.Polesya@cup.uni-muenchen.de}
\affiliation{%
  Department  Chemie,  Physikalische  Chemie,  Universit\"at  M\"unchen,
  Butenandstr.\ 5-13, 81377  M\"unchen, Germany\\}
\author{S.\ Mankovsky}
%\email{Sergiy.Mankovskyy@cup.uni-muenchen.de}
\affiliation{%
  Department  Chemie,  Physikalische  Chemie,  Universit\"at  M\"unchen,
  Butenandstr.\  5-13, 81377 M\"unchen, Germany\\}
\author{D.\ K\"odderitzsch} 
\affiliation{%
  Department  Chemie,  Physikalische  Chemie,  Universit\"at  M\"unchen,
  Butenandstr.\ 5-13, 81377  M\"unchen, Germany\\}
\author{J.~Min\'ar} 
\affiliation{%
  Department  Chemie,  Physikalische  Chemie,  Universit\"at  M\"unchen,
  Butenandstr.\ 5-13, 81377  M\"unchen, Germany\\}
\author{H.\ Ebert} 
\affiliation{%
  Department  Chemie,  Physikalische  Chemie,  Universit\"at  M\"unchen,
  Butenandstr.\ 5-13, 81377  M\"unchen, Germany\\}

\date{\today}
             
\begin{abstract}
  The temperature dependent stability of the magnetic phases of FeRh
  were investigated by means of total energy calculations with 
  magnetic disorder treated within the uncompensated
  disordered local moment (uDLM) approach. In addition, Monte Carlo
  simulations based on the extended Heisenberg model have been
  performed, using exchange coupling parameters obtained from first
  principles.
  The crucial role and interplay of two factors in the metamagnetic
  transition in FeRh has been revealed, namely the dependence of the
  Fe-Fe exchange coupling parameters on the temperature-governed degree
  of magnetic disorder in the system and the stabilizing nature of the
  induced magnetic moment on Rh-sites.  An important observation is the
  temperature dependence of these two competing factors.
\end{abstract}

\pacs{Valid PACS appear here}% PACS, the Physics and Astronomy
\maketitle

\section{Introduction}

FeRh with composition close to being equiatomic crystallizes in the CsCl
structure and exhibits rather interesting magnetic properties attractive
for investigation for various reasons. It is antiferromagnetically (AFM)
ordered in the ground state and reveals a first-order metamagnetic
transition to the ferromagnetic (FM) state at $T_m \approx 340-350$~K
\cite{KH62}. A transition to a paramagnetic (PM) state occurs at $T_\subel{C} =
675$~K \cite{KH62}. It is worth noting that the temperature of the 
metamagnetic transition is very sensitive to the conditions of sample
preparation. 
% which can be driven either by temperature of magnetic
% field. 
Despite several attempts to shed light on the physical origin of the magnetic
properties of FeRh-based alloys, both within experimental
\cite{MTF05,STK+08,MPH+13} and theoretical
\cite{GHE03,GA05,SM11,Der12, MM92, Mry05, BC14,SBD+14,KDT15} investigations, they are
still under debate.

Various mechanism have been suggested to explain the driving force for 
the AFM-FM transition. The early model  suggested by Kittel \cite{Kit60}, namely the
exchange-interaction-inversion model, associated the
AFM-FM phase transition with the dominant role of the change of
magnetoelastic energy. More recent investigations, however, demonstrated 
a minor role of the exchange magnetoelastic energy\cite{Kou66}. 

The experimentally observed large change of the entropy, i.e. 14.0 mJ/g/K
\cite{Kou66} or 12.58  mJ/g/K \cite{ANT+96}, is mainly attributed to
the electronic contribution related to 
spin fluctuations on the Rh atoms. This observation implies a
key role of the Rh induced magnetic moments for the stabilization of the FM
state \cite{THKC69,MML70,ANT+96}. This idea was supported by theoretical
investigations based on first-principles electronic structure
calculations \cite{GHE03,STK+08,SM11,  GA05,Mry05, SBD+14}.

So far there is no clear understanding
of the finite temperature magnetic properties of FeRh when 
small amounts of impurities are present \cite{Wal64,Kou66,BJL13,LNS09}.
Recent first-principles investigations by Staunton \emph{et al.} \cite{SBD+14},
based on the analysis of the electronic entropy
highlights the  impact of small compositional
changes on the temperature of the metamagnetic transition.

The reversibility of this transition and the small relaxation time
makes it attractive for applications.
Transport measurements demonstrate a strong drop of the electrical
resistivity during the metamagnetic phase transition \cite{BB95,ANT+96}
from the AFM to the FM state. As the metamagnetic transition can be
manipulated by an external magnetic field, this feature of the resistivity
leads to a giant magnetoresistance phenomena in the system around $T_m$,
that makes FeRh an appealing material for future data storage devices \cite{KMA01,KDT15}.

\section{Computational details}
\label{SEC:Computational-scheme}

Within the present study spin-polarized  electronic
structure calculations have been performed using the fully relativistic multiple scattering
KKR (Korringa-Kohn-Rostoker) Green function method  \cite{SPR-KKR6.3,EKM11}.
 All calculations have been performed in full-potential (FP) mode. 
Density functional theory employing the Generalized Gradient Approximation (GGA) 
was used with the parametrization of the exchange-correlation potential
as given by Perdew, Burke, Ernzerhof (PBE) \cite{PBE96}. For the angular
momentum expansion of the Green function a cutoff of
$\ell_{\mbox{\tiny max}} = 3$ was applied. For determining configurational averages in
substitutionally disordered alloys, the self-consistent coherent
potential approximation (CPA) method was employed. 
 The magnetic disorder in the DFT calculations for systems in the
paramagnetic (PM) state ($T > T_{\mbox{\tiny C}}$) was treated within the Disordered
Local Moment (DLM) method. To simulate the temperature induced partial
magnetic disorder below the Curie temperature, the so-called uncompensated
Disordered Local Moment (uDLM) approximation (see, e.g. Ref. \cite{KDTW08}) was used.  
In this case the Fe subsystem was represented by the
pseudo-alloy  Fe$_{1-x}^{\uparrow}$Fe$_{x}^{\downarrow}$ with Fe sites occupied by 
two-components Fe$^{\uparrow}$ and Fe$^{\downarrow}$ with the opposite directions of
magnetic moment, 'up' and 'down' respectively, and $x$ varying in the
interval $[0.0,0.5]$.

The finite temperature magnetic properties have been investigated via
Monte Carlo (MC) simulations based on the extended Heisenberg 
model, using a standard Metropolis algorithm \cite{Bin97,LB00}.
The exchange coupling parameters $J_{ij}$ for these calculations were
obtained using the expression given by Lichtenstein \cite{LKG84,LKA+87}.

\section{Results}

%%\subsection{Electronic structure calculations}

\begin{figure}[h]
\includegraphics[width=0.4\textwidth,angle=0,clip]{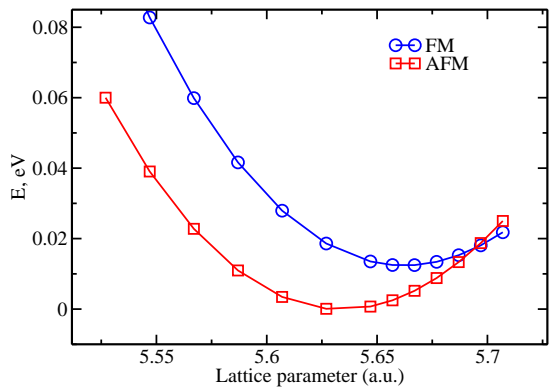}\;
\caption{\label{fig_ETOT_ALAT} Total energy calculated for the FM and
  AFM states of FeRh as a function of the lattice parameter $a$ }  
\end{figure}

The calculated total energies of  the FM and AFM states of FeRh as a
function of the  lattice parameter, represented in Fig. \ref{fig_ETOT_ALAT},
are in full agreement with the experimental results as well as with 
calculations of others authors \cite{DSB+14,GA05}. 
The energy minimum for the AFM state for the lattice parameter $a =
5.63$ a.u. is about $17$ meV lower than that for the FM state
occurring at  $a = 5.66$ a.u.
Thus, the AFM-FM transition should be accompanied by a lattice
expansion with a  magnetovolume effect of $~1.6\; \%$ which  
slightly overestimates the value observed experimentally, $~1\; \%$ \cite{MML70}.
%%%%%%%%%%%%%%%%%%%%%%%%%%%%%%%%%%%%%%%%%%%%%%%%%%%%%%%%%%%%%%%%%%%%%%%%%%%%%%%
%%%%%%%%%%%%%%%%%%%%%%%%%%%%%%%%%%%%%%%%%%%%%%%%%%%%%%%%%%%%%%%%%%%%%%%%%%%%%%%

%\subsubsection{Magnetic moments}

The Fe magnetic moment obtained for the energy
minimum in the FM state is $3.3\; \mu_B$, which is in line with the
neutron scattering measurements giving $3.0\; \mu_B$ per Fe atom
\cite{SNC64}. Its magnitude remains almost unchanged in the AFM ground
state ($3.2\; \mu_B$). 
The induced total magnetic moment on the  Rh atomic sites calculated for the FM state
is equal to $1.0\; \mu_B$ and vanishes in the AFM state.
%, that bring to
%discussions in the literature about its responsibility for stabilization  
%of the FM state \cite{}.

\begin{figure}[h]
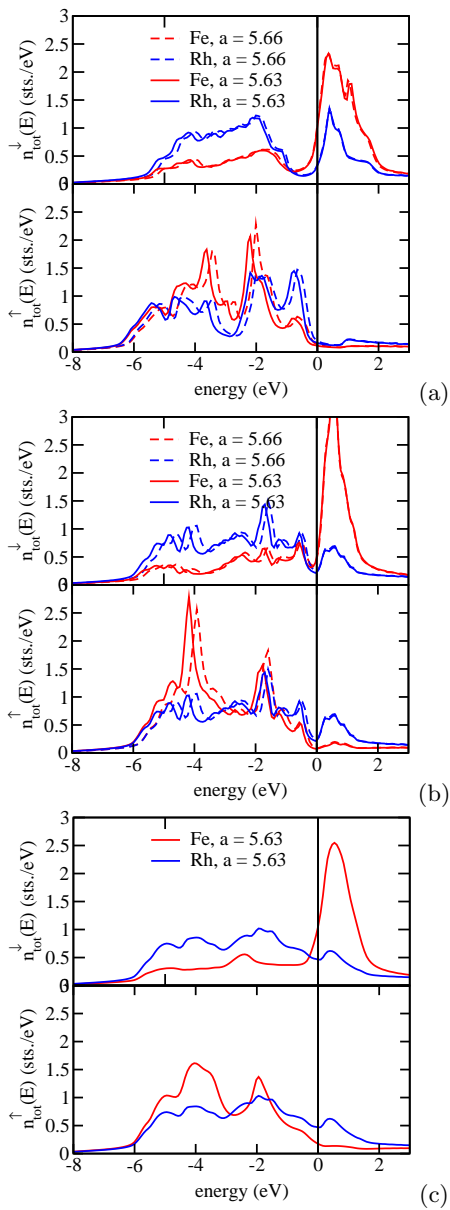

\includegraphics[width=0.3\textwidth,angle=0,clip]{CMP_FeRh_DOS.FM_vs_alat.eps}\;(a) \\
\includegraphics[width=0.3\textwidth,angle=0,clip]{CMP_FeRh_DOS.AFM_vs_alat.eps}\;(b) \\
\includegraphics[width=0.3\textwidth,angle=0,clip]{DLM_FeRh_DOS.dos.eps}\;(c)
\caption{\label{fig:DOS} Density of states for the FM state (a), AFM
  state (b) and random-spin state treated within the DLM approach
  (c). The solid and dashed lines in (a) and (b) represent the
  results for FeRh with the lattice parameters corresponding to the
  minimum of total energy for the AFM state (solid line, $a =
  5.63$~a.u.) and AFM state (dashed line, $a = 5.66$~a.u.)
 }  
\end{figure}

The density of states (DOS) curves for the FM, AFM and DLM 
states are shown in 
Fig. \ref{fig:DOS} (a), (b), (c), respectively. 
Only a weak dependency on the volume is observable for the DOS calculated 
for the FM (a) and AFM (b) states. In the FM state a strong spin-dependent
hybridization of Rh states is apparent (\ref{fig:DOS} (a)), which
leads to the formation  of a magnetic moment of $1\; \mu_B$ on the Rh atomic site. 
In the AFM state the Rh DOS for the majority- and minority-spin
states are identical due to symmetry, resulting in a Rh magnetic moment equal to $0\;
\mu_B$. The Rh-related electronic states for both of the DOS spin channels
in these cases exhibits hybridization with Fe minority-spin states
(essentially, above the Fermi energy) and majority-spin states (below
the Fermi energy), in line with the discussion by Sandratskii and
Mavropoulos \cite{SM11}.  This demonstrates that in the AFM state 
the vanishing total magnetic moment on the  Rh site is a result of 
the hybridization-governed redistribution of  spin density and not because 
of vanishing spin density  within the Rh atomic site. The DOS

Note also, that the
Rh DOS at the Fermi level is rather small and, therefore, a pronounced
Stoner enhancement of the  magnetic moment induced on   Rh can be ruled out. 
Thus, as will be also discussed below, the large Rh magnetic moment in the FM state
should be attributed to a strong spin-dependent hybridization.       
 
%%%%%%%%%%%%%%%%%%%%%%%%%%%%%%%%%%%%%%%%%%%%%%%%%%%%%%%%%%%%%%%%%%%%%%%%%%%%%%%
%%%%%%%%%%%%%%%%%%%%%%%%%%%%%%%%%%%%%%%%%%%%%%%%%%%%%%%%%%%%%%%%%%%%%%%%%%%%%%%

As the AFM-FM transition occurs at finite temperature it is necessary
 to take into account the temperature induced magnetic disorder in the
system when comparing the total energies of the two states.
For this reason the calculations have been performed accounting for
magnetic disorder treated within the uDLM approximation. 
In the case of the FM state with partial magnetic disorder, 
the normalized magnetic moment at each Fe site is 
$\bar{m}_\subel{Fe} = <M_\subel{Fe}>/M_\subel{Fe} =  (M_\subel{Fe}(1-x) -
M_\subel{Fe}x)/M_\subel{Fe} = (1-2x)$. 
In the case of the partially disordered AFM state, the same procedure
was applied to each Fe sublattice,
$M^{\uparrow}_\subel{Fe}$ and $M^{\downarrow}_\subel{Fe}$ having opposite alignment of
magnetic moments with respect to each other, i.e. $\bar{m}_\subel{Fe} =
<M^{\uparrow}_\subel{Fe}>/M^{\uparrow}_\subel{Fe} = <M^{\downarrow}_\subel{Fe}>/M^{\downarrow}_\subel{Fe} = (1-2x)$.
Thus, in this case the total magnetic moment is equal to $0\; \mu_\subel{B}$ for each
$\bar{m}_\subel{Fe}$ value. 

\begin{figure}[h]
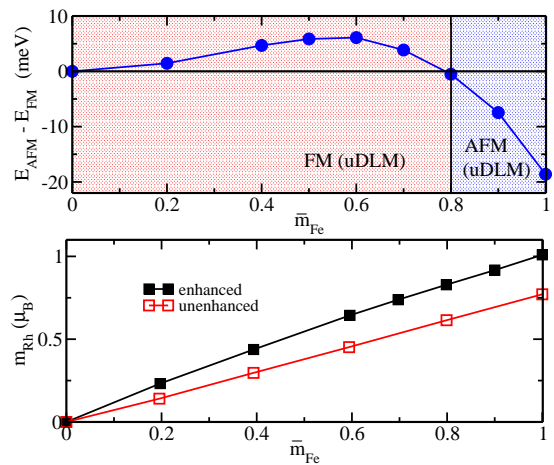

\includegraphics[width=0.4\textwidth,angle=0,clip]{CMP_FeRh_ETOT_NDLM_MRh_NDLM_a.eps}\;
\includegraphics[width=0.4\textwidth,angle=0,clip]{CMP_FeRh_ETOT_NDLM_MRh_NDLM_b.eps}\;
\caption{\label{fig_ETOT_NDLM} 
Top: Difference between the total energies, $E_\subel{AFM} -
E_\subel{FM}$ as a function of $\bar{m}_\subel{Fe}$, 
obtained by uDLM calculations 
for the FeRh compound having  FM and AFM magnetic order, with lattice
parameters corresponding to the total energy minimum in the AFM state, $a = 5.63$ a.u.
Bottom: Enhanced and non-enhanced magnetic moment on the Rh-site in the FM
(uDLM) state, as a function of average normalized magnetic moment on
Fe-sites, $\bar{m}_\subel{Fe}$.}
\end{figure}

Fig. \ref{fig_ETOT_NDLM} shows the total energy difference 
$E_\subel{AFM} - E_\subel{FM}$ as a function of $\bar{m}_\subel{Fe}$ for FeRh with
the lattice parameter $a = 5.63$ corresponding to the energy minimum of the
AFM state. At $x = 0$, the difference is negative, demonstrating the
stability of the AFM state in line with the results shown in Fig. \ref{fig_ETOT_ALAT}.
An increase of the disorder represented by a decrease of $(1-2x)$ leads to a decrease of
stability of the AFM state such that at $\bar{m}_\subel{Fe} \leq 0.8$ the FM state
becomes more stable up to the fully disordered state with $\bar{m}_\subel{Fe} = 1-2x
= 0.$, when both types of magnetic order, FM and AFM, have the same
energy.  The Rh magnetic moment in the case of FM order exhibits almost a
linear dependence on $\bar{m}_\subel{Fe}$ changing from $M_\subel{Rh} = 0.0\; \mu_B$
in the fully disordered DLM state to $M_\subel{Rh} = 1.0\; \mu_B$ in the ordered
FM state. 

In summary, these results demonstrate the following effects of  increasing
magnetic disorder: (i) a stabilization of the FM state with respect to the AFM state 
and (ii) the stability of the FM state is a result of the decrease 
of the Rh magnetic moment.

Therefore, when discussing the driving forces behind the  metamagnetic
AFM-FM phase transition, 
other additional effects than just the magnetization of the Rh
sublattice have to be considered. 
This has already become apparent  within various investigations
 \cite{GHE03,GA05,SM11,Der12, BC14,SBD+14,KDT15}. 
Below we will investigate the features of the interatomic exchange
interactions to demonstrate their crucial role for the AFM-FM phase
transition.

%%%%%%%%%%%%%%%%%%%%%%%%%%%%%%%%%%%%%%%%%%%%%%%%%%%%%%%%%%%%%%%%%%%%%%%%%%%%%%%
%%%%%%%%%%%%%%%%%%%%%%%%%%%%%%%%%%%%%%%%%%%%%%%%%%%%%%%%%%%%%%%%%%%%%%%%%%%%%%%

%%%\subsubsection{Magnetic moment induced on  Rh site}

First we discuss some features of the Rh magnetic moment which
are related to the magnetic disorder in the system.
The magnetic moment of Rh is induced by a spin dependent hybridization of 
its electronic states with the electronic states of surrounding Fe atoms.
This hybridization plays a crucial role  during the  transition from the FM to the AFM
state. To demonstrate the strong covalent character of the Rh magnetism, SCF 
calculations have been performed by suppressing for the spin-dependent part of the 
exchange-correlation potential ($B_\subel{xc}^\subel{Rh} = 0$) that is 
responsible for an enhancement of the spin magnetic moment induced
by the hybridisation with the Fe states.
Figure \ref{fig_ETOT_NDLM} (open symbols) shows
that the non-enhanced Rh magnetic moment is only $\sim$ 25 \% smaller than the
proper one. The same is observed  for the total energy. This demonstrates
the significant role of the spin dependent hybridization for the formation of
a large magnetic moment on the Rh site. As a result, the varying magnetic disorder
in the Fe sublattice in the presence of the  weak Rh exchange enhancement leads to
an almost linear change of $M_\subel{Rh}$ as a function of $\bar{m}_\subel{Fe}$.

%%%\subsubsection{Exchange coupling parameters}

\begin{figure}[h]
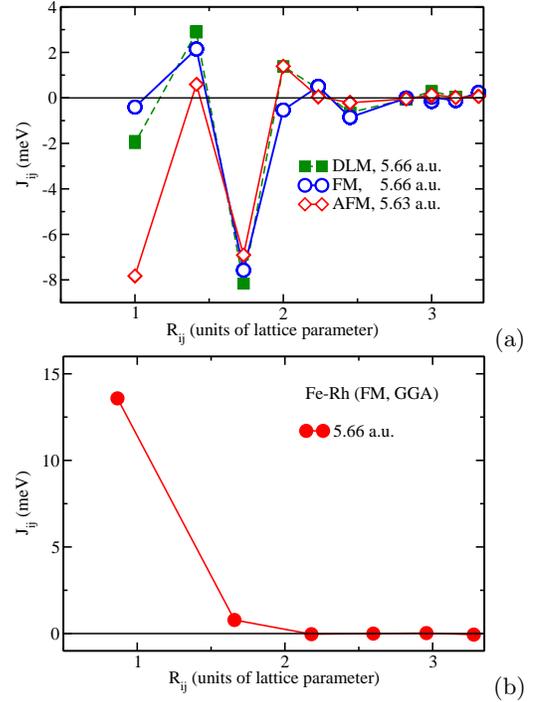

\includegraphics[width=0.35\textwidth,angle=0,clip]{CMP_Fe_Fe_FM_vs_AFM.eps}\;(a)
\includegraphics[width=0.35\textwidth,angle=0,clip]{Fe_Rh_FM.eps}\;(b)
\caption{\label{fig:JXC_AFM_UDLM} (a) The Fe-Fe interatomic exchange coupling parameters
calculated for the FM (open circles), AFM (open diamonds) and DLM
(squares) states; (b) the Fe-Rh exchange coupling parameters 
calculated for the FM state (the Rh magnetic moment is $m \approx 1\mu_B$)
 }  
\end{figure}

To investigate the stability of the FM and AFM ordered magnetic states at finite
temperature the exchange coupling parameters $J_{ij}$ have been
calculated for different reference states: AFM, FM and DLM.
These interactions can be seen to map the magnetic energy of
the system onto the Heisenberg Hamiltonian accounting for the bilinear
interatomic exchange terms.
The corresponding results are presented in Fig. \ref{fig:JXC_AFM_UDLM}.

Discussing these results, it is convenient to distinguish between the two Fe sublattices with
opposite directions of the magnetic moments in the AFM state. For
each Fe atom its first and third neighbor in the Fe subsystem 
belongs to another sublattice. One can see that for all reference states
the exchange couplings with these neighbours are negative indicating the
trend towards the formation of  AFM order. The interaction
with the third Fe neighbor depends only weakly on the reference state,
while the interactions with the first neighbour are close to $0$~meV in
the case of FM reference state and is about $-8.0$~ meV for the AFM state.
Since the Rh magnetic moment in the AFM state is equal to $0\; \mu_B$, it does
not contribute to the magnetic energy. As a consequence only the Fe-Fe exchange
interactions are responsible for the stabilization of this state.
In contrast to this situation, the FM order in the system can be stabilized by a rather
strong Fe-Rh interaction since the Rh magnetic moments are
non-zero, giving a negative contribution to the magnetic energy
competing with the positive one due to the Fe-Fe interatomic exchange. 
Thus, the transition from the AFM to FM state is essentially a 
result of the competition of these interactions. 

Thus, the behaviour of the FM and AFM energy variation shown in
Fig. \ref{fig_ETOT_NDLM} and demonstrating the stabilization of the FM state
upon increase of the magnetic disorder, can be attributed to the
modification of the Fe-Fe exchange coupling parameters, in particular,
to a strong decrease of the AFM interactions with the first neighbors. 
In the DLM state the situation is very different. The Fe-Fe exchange
interactions are rather close to those obtained for the FM state, but
the Rh magnetic moments are equal to zero and therefore give no contribution
to the magnetic energy.

%%%%%%%%%%%%%%%%%%%%%%%%%%%%%%%%%%%%%%%%%%%%%%%%%%%%%%%%%%%%%%%%%%%%%%%%%%%%%%%
%%%%%%%%%%%%%%%%%%%%%%%%%%%%%%%%%%%%%%%%%%%%%%%%%%%%%%%%%%%%%%%%%%%%%%%%%%%%%%%

%\subsection{Evaluation of $T_\subel{C}$}
%\label{meth-Ac}

%%%\subsection{Monte Carlo simulations based on extended Heisenberg model}
%%%\label{meth-Ac}

Two remarks concerning the approximations used for the calculations are in due place  
concerning the  Monte Carlo simulations.
The conventional Heisenberg Hamiltonian should be generalized  beyond the
classical form: (i) in order to account for the contribution due to induced
Rh magnetic moments; (ii) to account for not only bi-linear terms of the
magnetic energy expansion but also for terms of higher order.
The second generalization is required since the insufficient conventional dipolar
form of the Hamiltonian gets appreciable corrections from
bi-linear exchange coupling parameters calculated for different reference
states. The general form of the expansion of magnetic energy around a
reference state may therefor be represented as follows (see, e.g. \cite{MKS97,SKMR05,SKSM06})
%eeeeeeeeeeeeeeeeeeeeeeeeeeeeeeeeeeeeeeeeeeeeeeeeeeeeeeeeeeeeeeeeeeeeeee
%\cdot
\begin{eqnarray}
\label{Heisenberg_base}
E &=& E_\subel{ref} + \Delta E({|M_i|}) - \sum_{ij}J^{(2)}_{ij}(\hat{\vec{e}}_i \cdot
\hat{\vec{e}}_j) \\  \nonumber
& & - \sum_{\nu=2}^n\sum_{ij}J^{(2),(\nu)}_{ij}(\hat{\vec{e}}_i \cdot
\hat{\vec{e}}_j)^\nu \\ \nonumber
& & -\frac{1}{4!} \sum_{ijkl}J^{(4)}_{ijkl}[(\hat{\vec{e}}_i \cdot
\hat{\vec{e}}_j)(\hat{\vec{e}}_k \cdot \hat{\vec{e}}_l) + \\ \nonumber
& & (\hat{\vec{e}}_j \cdot \hat{\vec{e}}_k)(\hat{\vec{e}}_l \cdot \hat{\vec{e}}_i) +
(\hat{\vec{e}}_l \cdot \hat{\vec{e}}_i)(\hat{\vec{e}}_j \cdot \hat{\vec{e}}_k)] - ...
\end{eqnarray}
%
%eeeeeeeeeeeeeeeeeeeeeeeeeeeeeeeeeeeeeeeeeeeeeeeeeeeeeeeeeeeeeeeeeeeeeee
where $\Delta E({|M_i|})$ is the change in energy due to the change of
absolute values of local spin magnetic moments. This
can be reduced to the conventional form of the Hamiltonian with
redefined bilinear exchange interaction parameter $\tilde{J}$

%eeeeeeeeeeeeeeeeeeeeeeeeeeeeeeeeeeeeeeeeeeeeeeeeeeeeeeeeeeeeeeeeeeeeeee
%\cdot
\begin{eqnarray}
\label{Heisenberg_gen1}
H_\subel{ext} &=& -\sum_{\subel{Fe}: i,j }\left[{\tilde{J}^\subel{Fe-Fe}_{ij}(\bar{m})}
 +\sum_{\subel{Rh}: k}\tilde{J}^\subel{Fe-Rh}_{ik}{\chi_{kj}}\right](\vec{M}_i\cdot\vec{M}_j)  
\, .
\end{eqnarray}
%
%eeeeeeeeeeeeeeeeeeeeeeeeeeeeeeeeeeeeeeeeeeeeeeeeeeeeeeeeeeeeeeeeeeeeeee
The first term in Eq.~(\ref{Heisenberg_gen1}) 
characterizes the Fe-Fe transverse-fluctuation exchange energy with $i,j$ 
indicating sites on the Fe-sublattice   and in the
generalized form become dependent on the average magnetic moment in the
system. The second term describes the energy
changes related to longitudinal spin fluctuations on the Rh atoms
\cite{Mry05,PMS+10} with $k$ numbering Rh sublattice sites. 
The magnetic moments on the Fe site are denoted as $\vec{M}_{i(j)}$. 
Considering the DLM state as reference state the
dependence on the average magnetic moment in the first term can be
considered in linear approximation to have the following form
%eeeeeeeeeeeeeeeeeeeeeeeeeeeeeeeeeeeeeeeeeeeeeeeeeeeeeeeeeeeeeeeeeeeeeee
%\cdot
\begin{eqnarray}
\label{Heisenberg_gen}
\tilde{J}^\subel{Fe-Fe}_{ij}(\bar{m}) &=&  \tilde{J}^\subel{DLM}_{ij} 
+ \left[ \tilde{J}^\subel{FM/AFM}_{ij} - \tilde{J}^\subel{DLM}_{ij}\right] \bar{m}
\, .
\end{eqnarray}
%
%eeeeeeeeeeeeeeeeeeeeeeeeeeeeeeeeeeeeeeeeeeeeeeeeeeeeeeeeeeeeeeeeeeeeeee
The term characterizing the longitudinal contribution was discussed
previously  \cite{PMS+10}.
 
The response function $\chi_{kj}$ occuring in Eq.~(\ref{Heisenberg_gen1}) 
describes the Rh magnetic moment induced by surrounding Fe atoms and is dependent
on the orientation of their magnetic moments. A linear approximation expressed by a constant
$\chi_{kj}$ was used, that is based on the results above
showing the almost linear dependence of induced Rh magnetic moment on
the average magnetic moment in Fe subsystem (see Fig. \ref{fig_ETOT_NDLM}). 
As the Rh magnetic moment occurs essentially due to the spin-dependent
hybridization of the Rh electronic states with the electronic states of
neighboring Fe atoms, it is represented in MC simulations through the
average magnetic moment on the first Fe neighbor shell around the Rh 
atoms, leading to an approximate form for the susceptibility function
\cite{PMS+10} 
\begin{eqnarray}
\label{A3}
\vec{m}^\subel{Rh} &=& \sum_{j} \chi^\subel{Rh-Fe}_{0j}\vec{M}_j =
X^\subel{Rh-Fe} \sum_{j} \vec{M}_j \;,
\end{eqnarray}
where the summation is performed over the magnetic moments $\vec{M}_{j}$ 
corresponding to Fe atoms  within the
first-neighbor shell around the 'non-magnetic' Rh atom on site $i=0$.

It should be mentioned in addition that the same DLM reference state for both the
re-scaled FM and AFM exchange interactions has been used. This means that the exchange
interactions should change abruptly at the metamagnetic transition point
that accounts for latent heat connected with the first-order phase
transition.

%%%\subsection{Finite-temperature magnetism}

%%%%%%%%%%%%%%%%%%%%%%%%%%%%%%%%%%%%%%%%%%%%%%%%%%%%%%%%%%%%%%%%%%%%%%%%%%%%%%%
\begin{figure}[h]
\includegraphics[width=0.35\textwidth,angle=0,clip]{Ni-FeRh_MC_M_vs_T_vs_experiment.eps}\;(a)
\includegraphics[width=0.35\textwidth,angle=0,clip]{CMP_MOM_vs_experiment_fe_impur.eps}\;(b)
\includegraphics[width=0.2\textwidth,angle=0,clip]{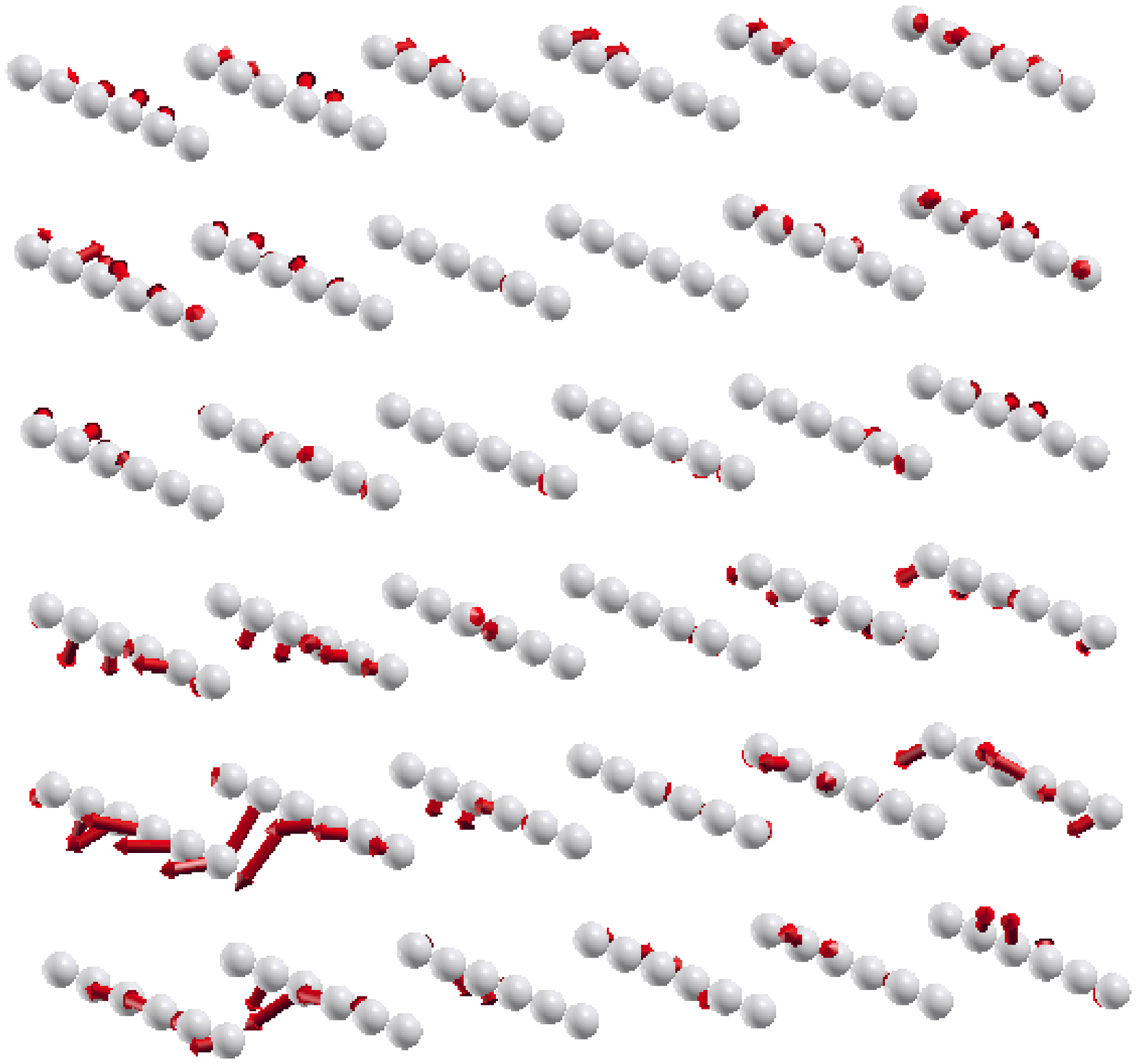}\;(c)
\includegraphics[width=0.2\textwidth,angle=0,clip]{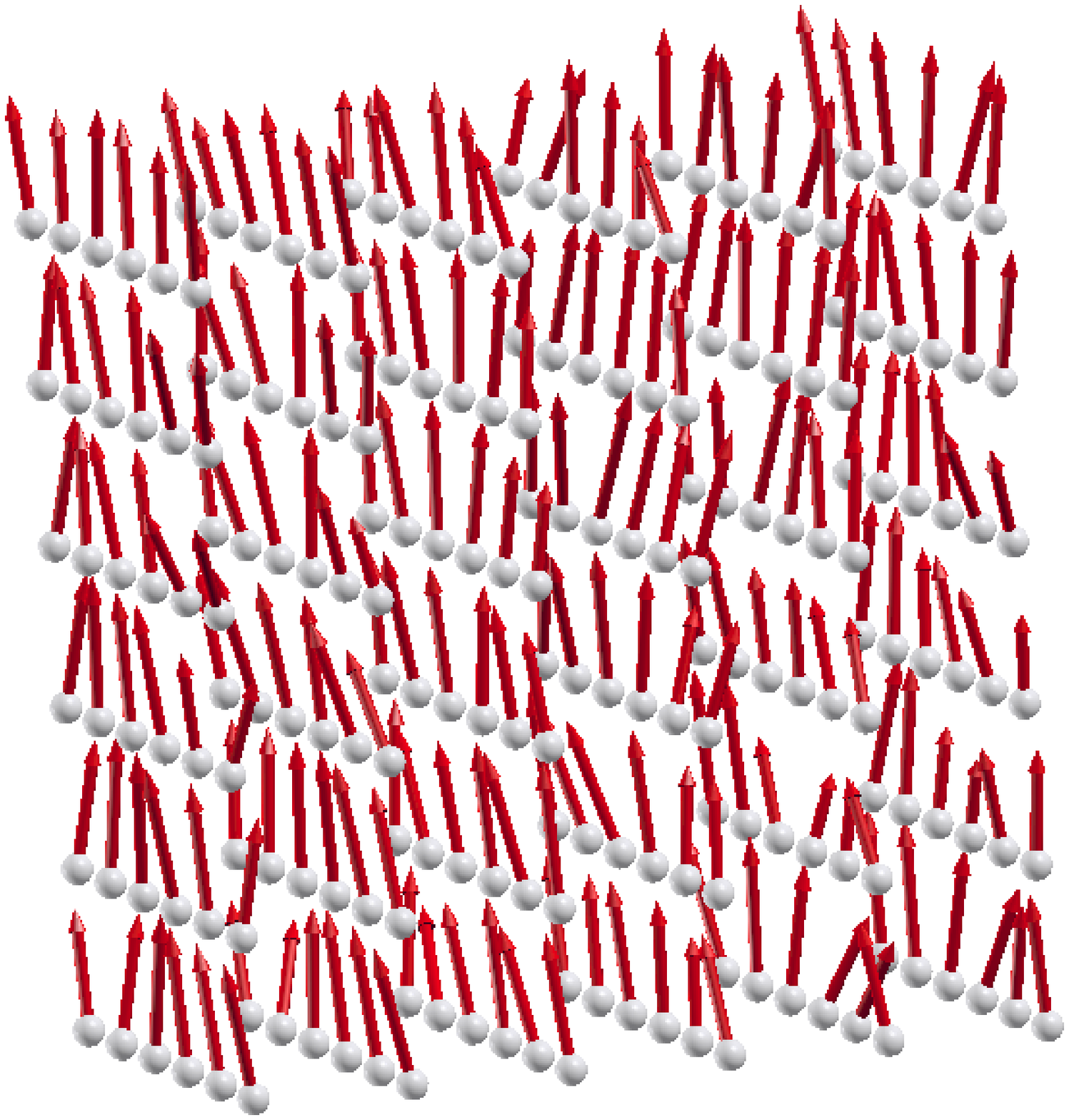}\;(d)
\caption{\label{fig:M_MC_vs_Expt} 
Temperature dependent relative magnetization $M(T)/M_0$ ($M_0$ is the
magnetization at $T = 0$~K) obtained within the
MC simulations: (a) $M(T)/M_0$ for pure FeRh (circles) in comparison with
the experimental result; squares and diamonds represents the results for
FeRh with 5 (diamonds) and 10 $\%$ (squares) substitution of Fe by  Ni
atoms; (b) $M(T)/M_0$ calculated for FeRh with 1 and 2 $\%$ of Fe (closed symbols) in the Rh
sublattice, in comparison with the results for FeRh (open symbols). (c)
and (d) show the induced magnetic moments on the Rh sublattice of
pure FeRh at $T = 200$ and $340$ K, respectively.}  
\end{figure}
%%%%%%%%%%%%%%%%%%%%%%%%%%%%%%%%%%%%%%%%%%%%%%%%%%%%%%%%%%%%%%%%%%%%%%%%%%%%%%%

To investigate the finite temperature magnetic properties of FeRh-based
systems, the Monte
Carlo (MC) simulations have been performed based of the
model Heisenberg Hamiltonian, Eq.~(\ref{Heisenberg_gen1}). 
In a first step the calculations have been performed for the pure FeRh compound.
The  magnitude of the  Fe magnetic moments have been fixed and only 
changes in orientation have been considered. On the other hand, the magnetic moments
treated as induced magnetic moment according to Eq. (\ref{A3}),
change their absolute value as well as the orientation depending on the 
orientations of the magnetic moments of the surrounding Fe atoms. 
At the same time, the total magnetic moment in the system can be rather
small approaching $0\; \mu_B$ at low (AFM state) and high (PM state)
temperatures.   
Figure \ref{fig:M_MC_vs_Expt} shows the relative magnetization as a function
of the temperature in comparison to experimental results. The calculated AFM-FM 
transition occurs at $T = 320$~K, rather close to the experimental
value $T = 350$~K. As it was discussed above, it is caused by the
increasing magnetic disorder in the system when the temperature 
increases. Two mechanisms are the major driving force for the transition.
Firstly, the 
disorder-induced modification of the exchange coupling parameters. Secondly,
the increase of the amplitude of randomly oriented fluctuations of the Rh
magnetic moments in the AFM state due to increasing temperature-induced
short-range FM order in the Fe subsystem (see Fig. \ref{fig:M_MC_vs_Expt}(c)).
Therefore,
it is the  occurrence of magnetic moments on the Rh sites above a
certain temperature that leads to a stabilization of the FM order in the
system (see Fig. \ref{fig:M_MC_vs_Expt}(d)). A further temperature
increase results in a decrease of the Rh 
magnetic moment, and to a transition to the PM state at $T = 720$~K.   
One has to stress the asymmetry of the metamagnetic transition in FeRh
upon heating and cooling of the sample. This was demonstrated recently
by a robust experimental investigation on the formation of FM and AFM phases.
The authors concluded that the formation of the AFM phase upon a temperature
decrease is dominated by a nucleation at defects in contrast to the
formation of the FM phase for increasing temperature due to heterogeneous 
nucleation at different sites \cite{BBA+15}. The latter results are in line with the
 present results of MC simulations (e.g. see
 Fig. \ref{fig:M_MC_vs_Expt}(c),(d)). However, different mechanism of
 nucleation upon cooling requires further generalization of the
 Hamiltonian and more sophysticated spin dynamics simulations 
to reproduce the temperature dependent behaviour of
 magnetization in this case.

To investigate the influence of impurities on the metamagnetic
transition, the calculations have been performed for the FeRh systems
with $5$ at.\% and $10$ at.\% substitution of Fe by Ni atoms, 
and with $1$ at.\% and $2$ at.\% substitution of Rh by Fe atoms.

\begin{figure}[h]
\includegraphics[width=0.35\textwidth,angle=0,clip]{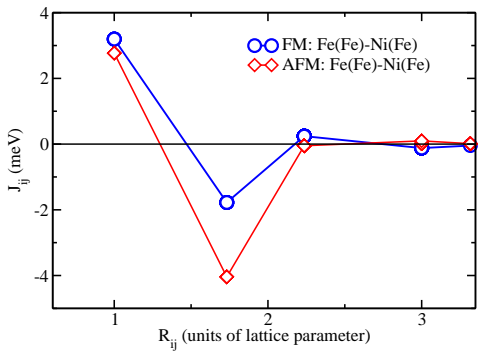}\;
\caption{\label{fig:FeNi_Rh} (a) The Fe-Ni interatomic exchange
  coupling parameters calculated for the FM (open circles) and AFM (open
  diamonds) FeRh compound with $5\; \%$ substitution of Fe by Ni atoms: The Fe
  and Ni atoms correspond to different Fe sublattices having in the AFM
  state an opposite orientation of the magnetization.
 }  
\end{figure}

The presence of Ni impurities results in a decrease of the temperature
of the metamagnetic transition. As one can see in
Fig.~\ref{fig:M_MC_vs_Expt} (a), 5 and 10 atomic percent of Ni in the Fe
sublattice leads to transition temperatures of  $T_m = 230$ and  $180$~K, respectively.
The decrease of $T_m$  is mainly governed by the difference in the Fe-Ni exchange
interactions when compared to the Fe-Fe exchange interactions. As one
can see in  Fig. \ref{fig:FeNi_Rh}, the Fe exchange interaction with the
Ni atom at the first-neighbor position becomes positive. The Fe-Ni exchange
interactions, when Ni occupies third-neighbor position, are negative but
are much smaller in magnitude when compared to the Fe-Fe interactions.
Both of these effects lead to a stabilization of the FM state, and as a
consequence to a decrease the temperature of metamagnetic AFM-FM transition.
%%%%%%%%%%%%%%%%%%%%%%%%%%%%%%%%%%%%%%%%%%%%%%%%%%%%%%%%%%%%%%%%%%%%%%%%%%%%%%%
%%%%%%%%%%%%%%%%%%%%%%%%%%%%%%%%%%%%%%%%%%%%%%%%%%%%%%%%%%%%%%%%%%%%%%%%%%%%%%%

\begin{figure}[h]
\includegraphics[width=0.35\textwidth,angle=0,clip]{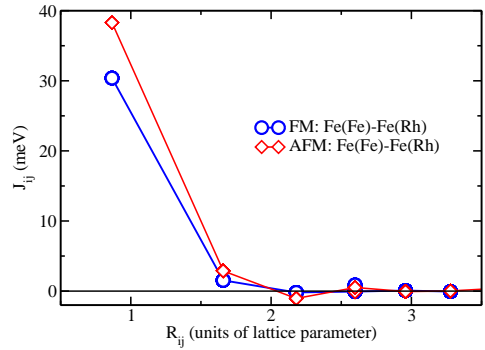}\;
\caption{\label{fig:Fe_RhFe} (a) The Fe-Fe interatomic exchange
  coupling parameters calculated for the FM (open circles) and AFM (open
  diamonds) FeRh compound with $1\; \%$ substitution of Rh by Fe atoms: 
  one Fe atom belong to the Fe sublattice, and another one to the Rh sublattice.
 }  
\end{figure}

Substitution of 1 and 2  atomic \% of Rh by Fe atoms results in
a decrease of the transition temperature from $T_m = 320$ to $T_m = 260$ and $220$~K, respectively.
In contrast to Fe substitution by Ni, the decrease of  $T_m$
is controlled by strong FM interactions between Fe atoms in the different
(Fe and Rh) sublattices (see Fig. \ref{fig:Fe_RhFe}).

Thus, in line with experiment, for both types of impurities we have
obtained a decrease of the temperature of the metamagnetic
transition. On the other hand, the effect of impurities is much weaker
than observed in experiment. This is clearly the result of
approximations used in our calculations, in particular, for the exchange
coupling parameters: (i) we use here the re-scaled bi-linear exchange
interactions in the model Hamiltonian, Eq.~(\ref{Heisenberg_gen1}); (ii) the first-principles
calculations of $J_{ij}$ are performed for the collinear magnetic state at
$T = 0$~K, which can be crucial for such a delicate system as FeRh. This
problem can be avoided for example by the self-consistent DLM approach by Staunton et
al. \cite{SBD+14}, that leads, however, to much more time-consuming
calculations of the temperature dependent properties.

%%%%%%%%%%%%%%%%%%%%%%%%%%%%%%%%%%%%%%%%%%%%%%%%%%%%%%%%%%%%%%%%%%%%%%%%%%%%%%%
%%%%%%%%%%%%%%%%%%%%%%%%%%%%%%%%%%%%%%%%%%%%%%%%%%%%%%%%%%%%%%%%%%%%%%%%%%%%%%%

\section{Summary}

To summarize, we have studied here the AFM-FM metamagnetic transition in
FeRh on the basis of the first-principles DFT calculations. The temperature
dependent stability of these phases was investigated performing 
total energy calculations for the systems with a different degree of
magnetic disorder treated within the uDLM approach. The
first-principles calculations supply in addition the parameters (element projected
magnetic moments, exchange coupling parameters) for the extended
Heisenberg model Eq.~(\ref{Heisenberg_gen1}).
Based on this Hamiltonian, Monte Carlo simulations have been
performed. The results of both calculations
allow to identify the crucial role and interplay of
two factors: (i) the dependence of the Fe-Fe exchange coupling parameters on
the temperature-governed degree of magnetic disorder in the system;
(ii) the Rh induced magnetic moment, also dependent on the magnetic disorder
in the system, that stabilize the FM state. An important observation is
the competing effect of the temperature dependence of these two
factors. Increase of disorder for rising temperature leads to a decrease
of the Rh magnetic moments and as a result to a decrease of
Fe-(Rh)-Fe FM exchange interactions responsible for lowering the
energy of the FM state. On the other hand, the decrease of the AFM Fe-Fe
$\tilde{J}^\subel{Fe-Fe}_{ij}(\bar{m})$ exchange 
interactions (see extended Hamiltonian Eq.~(\ref{Heisenberg_gen1})
together with the temperature induced magnetic disorder leads to a
stabilization of the FM state. Suppressing the interplay of these two
effects leads to a shift of the point of metamagnetic transition.
This was demonstrated by studing the impact of impurities either on the Fe
or on the Rh sublattices.

\section{Acknowledgements}

Financial support by the DFG via SFB 689 (Spinph\"anomene in reduzierten
Dimensionen) is thankfully acknowledged.

%\bibliographystyle{prsty}
%\bibliography{/opt/ak/bib/akhelit,curie-in-induced,H_HEISENBERG}

\end{document}